\documentclass[reprint,amsmath,amssymb,aps,pre]{revtex4-1}  % for review and submission
\usepackage{graphicx}% Include figure files
\usepackage{caption}
\usepackage{subcaption}
\usepackage{dcolumn}% Align table columns on decimal point
\usepackage{bm}% bold math
\usepackage{hyperref}% add hypertext capabilities
\hypersetup{colorlinks=True,citecolor=blue,linkcolor=blue,urlcolor=blue}
\usepackage{multirow} % for tables
\usepackage{enumerate}

\def\be{\begin{equation}} %Equations for short
\def\ee{\end{equation}} %Equations for short
\def\intd{\,\mathrm{d}} %The differential symbol

\begin{document}

\title{Formation and Evolution of Target Patterns in Cahn-Hilliard Flows}
\author{Xiang Fan}\affiliation{University of California at San Diego, La Jolla, California 92093}
\author{P. H. Diamond}\affiliation{University of California at San Diego, La Jolla, California 92093}
\author{L. Chac\'on}\affiliation{Los Alamos National Laboratory, Los Alamos, New Mexico 87545}
\date{\today} 

\begin{abstract}
We study the evolution of the concentration field in a single eddy in the 2D Cahn-Hilliard system to better understand scalar mixing processes in that system. This study extends investigations of the classic studies of flux expulsion in 2D MHD and homogenization of potential vorticity in 2D fluids. Simulation results show that there are three stages in the evolution: (A) formation of a `jelly roll' pattern, for which the concentration field is constant along spirals; (B) a change in isoconcentration contour topology; and (C) formation of a target pattern, for which the isoconcentration contours follow concentric annuli. In the final target pattern stage, the isoconcentration bands align with stream lines. The results indicate that the target pattern is a metastable state. Band merger process continues on a time scale exponentially long relative to the eddy turnover time. The band merger process resembles step merger in drift-ZF staircases; this is characteristic of the long-time evolution of phase separated patterns described by the Cahn-Hilliard equation.
\end{abstract}

\maketitle

%\section{Introduction}

Spinodal decomposition is a process by which a binary liquid mixture can evolve from a miscible phase (e.g., water+alcohol) to two co-existing phases (e.g., water+oil). When the binary liquid mixture is cooled below the critical temperature in the absence of external forcing, initially small blobs coalesce and become larger blobs until their size grows to the system size \cite{cahn_free_1958}. If large scale external forcing is imposed, blob size growth is arrested. The competition between the elastic energy and the turbulent kinetic energy leads to a statistically stable blob size. The Hinze scale $L_H\sim (\frac{\rho}{\sigma})^{-3/5}\epsilon^{-2/5}$ is an estimate of the stable blob size. Here, $\rho$ is density, $\sigma$  is surface tension, and $\epsilon$ is the energy dissipation rate per unit mass \cite{pal_binary-fluid_2016,perlekar_two-dimensional_2017,perlekar_droplet_2012,perlekar_spinodal_2014,datt_morphological_2015,tierra_numerical_2014,guillen-gonzalez_second_2014}. In our previous work \cite{fan_cascades_2016}, we defined the elastic range to be the scales in the range $L_d<l<L_H$, where $L_d$ is the dissipation scale. The elastic range scales are those for which the surface tension-induced elasticity is important to dynamics.

The Cahn-Hilliard equation is a standard model for spinodal decomposition. When considering the back reaction of the surface tension on to fluid motion, we need to couple the Cahn-Hilliard with Navier-Stokes (CHNS). 2D CHNS has analogies to 2D Magnetohydrodynamics (MHD) \cite{ruiz_turbulence_1981}. The concentration $\psi$ in 2D CHNS is the analogue of the magnetic potential $A$ in 2D MHD. Both models consist of a vorticity equation and a “diffusion” equation for an active scalar. 2D CHNS differs from 2D MHD by the appearance of negative diffusivity for potential and a nonlinear dissipative flux. A linear elastic wave, the analogue of the Alfven wave, exists in the 2D CHNS system and introduces the crucial element of memory. This wave propagates along the \textit{interface} of the blobs, thus resembles a capillary wave. The two systems have identical ideal quadratic conserved quantities, and they both exhibit dual cascades. Our previous work \cite{fan_cascades_2016} showed that the mean square concentration spectrum for the 2D CHNS system in the elastic range is $\sim k^{-7/3}$, and it is associated with an inverse cascade of mean square concentration. Note that the power $-7/3$ is the same as the power for the mean square magnetic potential spectrum in 2D MHD. On the other hand, the kinetic energy spectrum is proportional to $k^{-3}$, which is the same power law as for a 2D Navier-Stokes fluid in the forward enstrophy cascade regime. The kinetic energy power law $-3$ in 2D CHNS is far from $-3/2$ in 2D MHD, and the difference can be explained by the difference in the physics of back reaction. Unlike the case in MHD where the magnetic fields fill the whole space, the CHNS analogue of the magnetic fields, which are the blob interfaces, have a much smaller spatial packing fraction (i.e., relative spatial ``active volume"). Thus, in CHNS, the back reaction is only significant in the interfaces of the blobs, because the waves propagate along the interfaces, like surface waves. This implies that for CHNS, the interfaces of the blobs are crucial to the mixing dynamics.

In order to better understand the dynamics in the CHNS turbulence, we examine the evolution of the concentration field in the background of a single convective eddy in the Cahn-Hilliard system. Since the system tends to evolve to a state of a few large blobs, the simplest problem which emerges is that of understanding the competition of shearing and dissipation in the context of a single cell structure. This goal leads us to a study which re-visits the classic problems of flux expulsion in 2D MHD \cite{weiss_expulsion_1966, gilbert_flux_2016, moffatt_magnetic_1983,moffatt_time-scale_1983} and potential vorticity homogenization in 2D fluids \cite{rhines_homogenization_1982,rhines_how_1983}. Weiss (1966) studied the evolution of an initial uniform magnetic field $\mathbf{B}_0$ in the background of a single eddy in 2D MHD \cite{weiss_expulsion_1966}. Because the magnetic field was expelled to a layer at the boundary of the eddy, this phenomenon was named ``flux expulsion". The eddy was observed to stretch the initial field, and the final value of average magnetic field was
estimated to be $\langle B^2\rangle\sim\mathrm{Rm}^{1/2}\langle B_0^2\rangle$, where $\mathrm{Rm}$ is the magnetic Reynolds number ($\mathrm{Rm}\gg1$). The time scale for the magnetic field to reach a steady state was found to scale as $\tau_{\mathrm{MHD}}\sim\mathrm{Rm}^{1/3}\tau_0$. Rhines and Young (1983) studied the time scale of the homogenization of a passive scalar in 2D fluids with closed streamlines \cite{rhines_how_1983}. A rapid stage and a later slow stage are observed. In the rapid stage, shear-augmented diffusion dominates, and the initial values of the passive scalar approach an average about a streamline. The time scale for the rapid stage is proportional to $\mathrm{Pe}^{1/3}$, where $\mathrm{Pe}$ is the P\'eclet Number. In the slow stage, the passive scalar homogenizes within the eddy over the full diffusion time, which is proportional to $\mathrm{Pe}$. A system of a few blobs can be viewed as an array of such eddies. Thus, we hope that understanding the physics of a single eddy can promote understanding of the turbulent system.
 
%\section{Simulation System}

\begin{figure}[htbp] %  figure placement: here, top, bottom, or page
    \centering
    \includegraphics[width=\columnwidth]{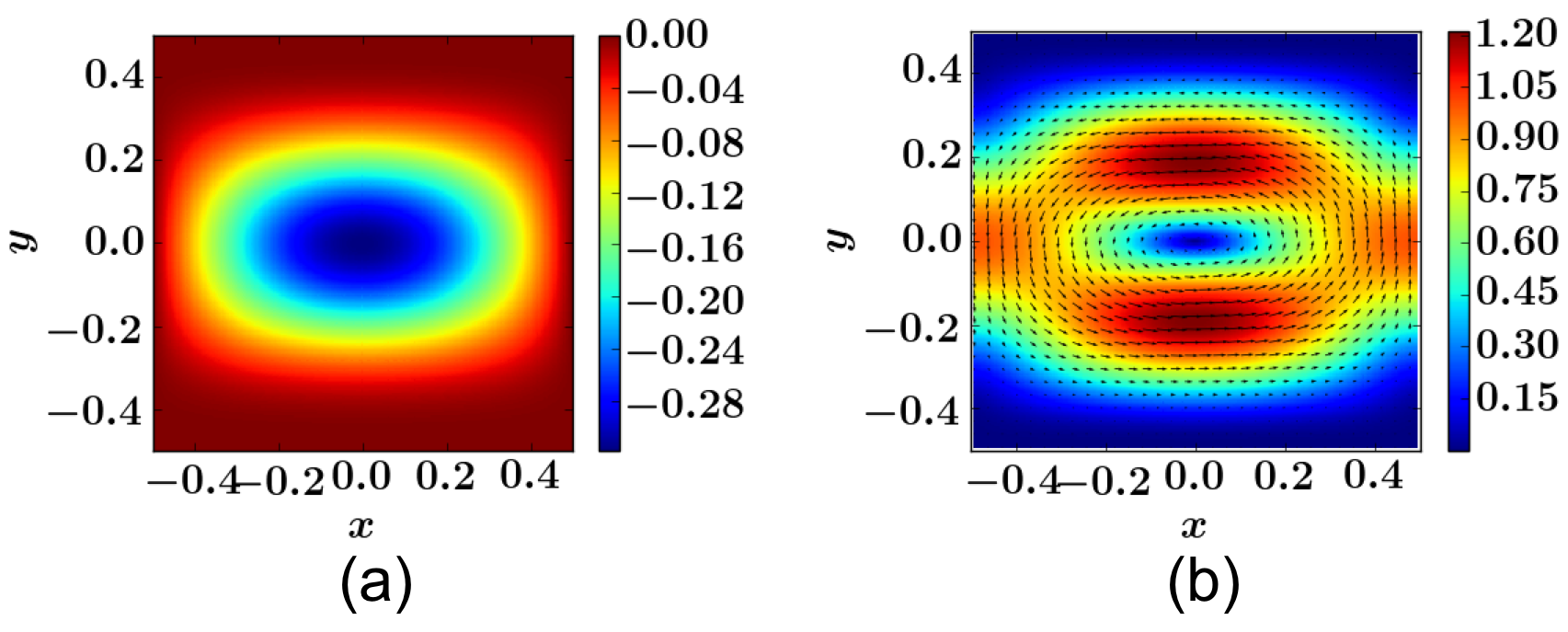}
    \caption{The background stream function $\phi$ (a) and velocity field $\mathbf{v}$ (b).}
    \label{background_velocity}
\end{figure}

In this Rapid Communication, we report on solutions of the 2D Cahn-Hilliard system by PIXIE2D \cite{chacon_implicit_2002,chacon_hall_2003} in the background of a single eddy, which is an analogue to the flux expulsion problem. The basic equation is:
\be
\partial_t\psi+\mathbf{v}\cdot\nabla\psi=D\nabla^2(-\psi+\psi^3-\xi^2\nabla^2\psi)
\ee
where $\psi=\frac{\rho_1-\rho_2}{\rho_1+\rho_2}$ is the local relative concentration and satisfies $-1\leq\psi\leq1$, $D$ is the diffusivity, $\xi$ is the parameter for interaction strength, and $\mathbf{v}$ is a background velocity field which does not change with time. We use the following background velocity field as a model of single eddy \cite{weiss_expulsion_1966}:
\be
\phi=-\frac{\phi_0}{\pi}(1-4y^2)^4\cos(\pi x)
\ee
where $\phi$ is the stream function, see Fig.~\ref{background_velocity}. Our simulation is defined on a box of size $L_0^2$ with $256^2$ points, $x,y\in[-L_0/2,L_0/2]$. The initial condition is $\psi_0(x,y)=B_0(x+L_0/2)$, where $B_0$ is a coefficient analogous to the magnitude of the external magnetic field in MHD. We set $B_0=1.0*10^{-2}$ in our runs to compare with Weiss's study \cite{weiss_expulsion_1966} ($B_0$ should be a small number considering the allowed range of $\psi$: $\psi\in[-1,1]$). The boundary conditions are Dirichlet in both directions: $\psi=\psi_0$ and $\nabla^2\psi=\nabla^2\psi_0$ at boundaries. Without losing generality, we normalize the system as follows: let length scale $L_0=1.0$ and time scale $t_0=L_0/v_0=L_0^2/\phi_0=1.0$. The absolute value of the velocity does not change the physics as long as it is nonzero, because we can always re-scale the time scale to make the system identical to the case where $v_0=1.0$. In this Letter, we focus on the competition between shearing and dissipation, so the $v_0=0$ cases are excluded. Thus, there are only two independent parameters: $D$ and $\xi$. The range of parameters used in our simulations are summarized in Table.~\ref{parameters}.

The dimensionless parameters are:
\begin{enumerate}
\item The P\'eclet Number $\mathrm{Pe}=L_0v_0/D$, the analogue of magnetic Reynolds number $\mathrm{Rm}$ in MHD, is the advective transport rate to the diffusive transport rate. In our simulation, $\mathrm{Pe}=D^{-1}$.
\item The Cahn Number $\mathrm{Ch}=\xi/L_0$ is the characteristic length scale of the interface width over the system size. In our simulation, $\mathrm{Ch}=\xi$. 
\end{enumerate}

\begin{table}[htbp]
\centering
\caption{The parameters used in our simulations.}
\label{parameters}
\begin{tabular}{cccc}
\hline
\hline
Runs   & $D$            & $\xi$  \\
\hline
Run1 & $3.16*10^{-4}$ & $1.0*10^{-2}$ \\
Run2 & $1.0*10^{-4}$  & $1.0*10^{-2}$ \\
Run3 & $3.16*10^{-5}$ & $1.0*10^{-2}$ \\
Run4 & $1.0*10^{-5}$  & $1.0*10^{-2}$ \\
Run5 & $3.16*10^{-5}$  & $1.2*10^{-2}$ \\
Run6 & $3.16*10^{-5}$ & $1.5*10^{-2}$ \\
Run7 & $3.16*10^{-5}$  & $1.8*10^{-2}$ \\
\hline         
\end{tabular}
\end{table}

%\section{Time Evolution of the Concentration Field}

\begin{figure*}[htbp]
   \centering
   \includegraphics[width=\textwidth]{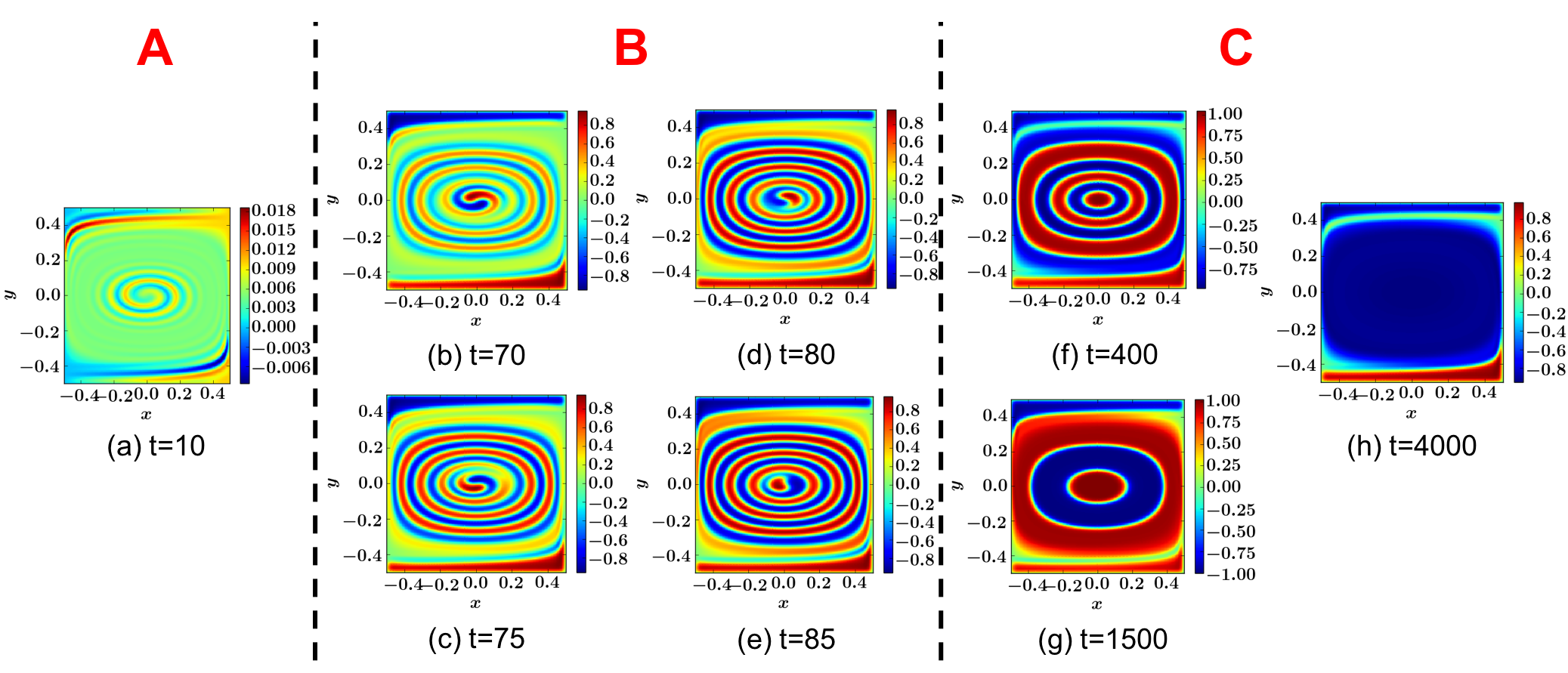} % requires the graphicx package
   \caption{The evolution of the $\psi$ field, represented by Run2. (a) The `jelly roll' stage, in which the stripes are spirals. (b) - (e) The topological evolution stage, in which the topology evolves from spirals to concentric annuli in the center of the pattern. (f) \& (g) The target pattern stage, in which the concentration field is composed of concentric annuli. The band merger progress occurs on exponentially long time scales; see (f) $\rightarrow$ (g) as an example. (h) The final steady state.}
   \label{psi_fields}
\end{figure*}

We observe three stages in the evolution of the concentration field: (A) formation of a `jelly roll' pattern, (B) a change in topology, and (C) formation of target pattern (Fig.~\ref{psi_fields}).

\begin{enumerate}[(A)] 
\item In the `jelly roll' pattern stage, the magnitude of $|\psi|$ remains close to zero. Stripes form gradually, and are then wound up into spirals by the fluid motion. See Fig.~\ref{psi_fields} (a) for a typical concentration field plot in the `jelly roll' pattern stage. This wind-up process also occurs in the early stage of the expulsion problem in MHD.

\item In the topological evolution stage, the $\psi$ field inside the stripes quickly approaches $\pm\sim 1$, demonstrating that phase separation has occurred. In this stage, stripes break up and reconnect with adjacent stripes. The spirals evolve to concentric annuli, with topology change proceeding from outside to inside, one annulus at a time. Fig.~\ref{psi_fields} (b) - (e) show the topology evolution of the stripe in the center of the pattern. Fig.~\ref{topology_change_cartoon} illustrates this process: the stripes break in the middle, while the outer parts reconnect to form a circle.

\begin{figure}[htbp]
   \centering
   \includegraphics[width=\columnwidth]{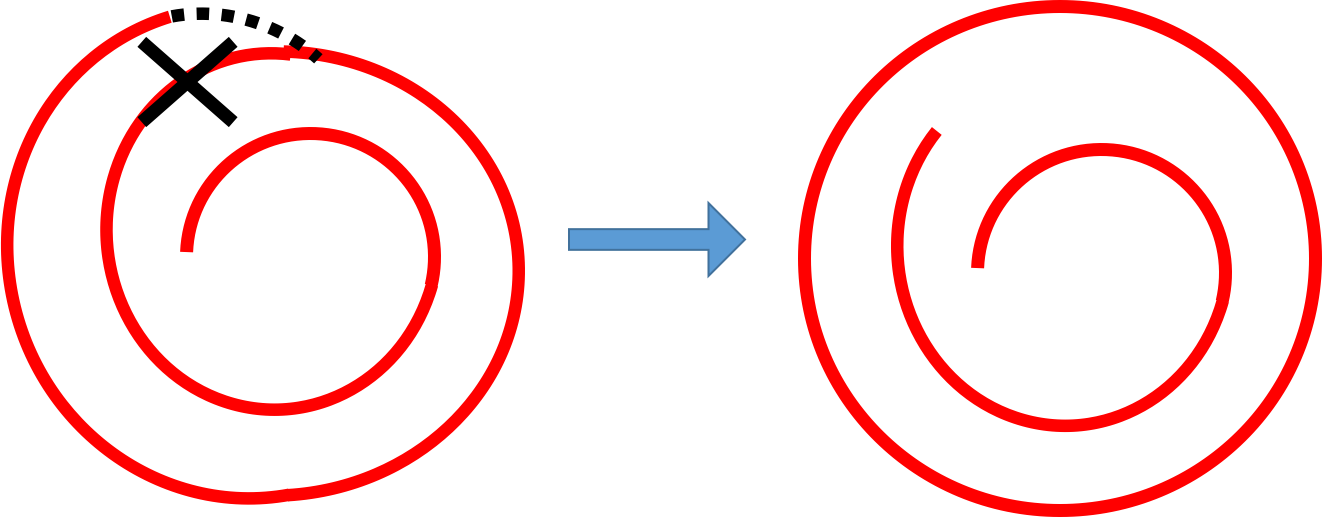} % requires the graphicx package
   \caption{An illustration of the topological evolution from the `jelly roll' pattern to the target pattern: the stripes break in the middle, and the outer parts reconnect into a circle.}
   \label{topology_change_cartoon}
\end{figure}

\item In the target pattern stage, the bands form concentric annuli, in contrast to spirals. See Fig.~\ref{psi_fields} (f) and (g) for typical concentration field plots in the target pattern stage. The bands are aligned with the stream lines. This structure is caused by shear flows. Shears can stabilize the bands against their intrinsic surface tension instabilities. Previous studies that investigated how the Cahn-Hilliard system behaves in a shear flow noted that the formation of band patterns aligned along the flow direction \cite{liu_isogeometric_2013,hashimoto_phase_1993,hashimoto_string_1995}. Shear flow with closed stream lines leads to the target pattern.
\end{enumerate}

\begin{figure}[htbp]
   \centering
   \includegraphics[width=0.8\columnwidth]{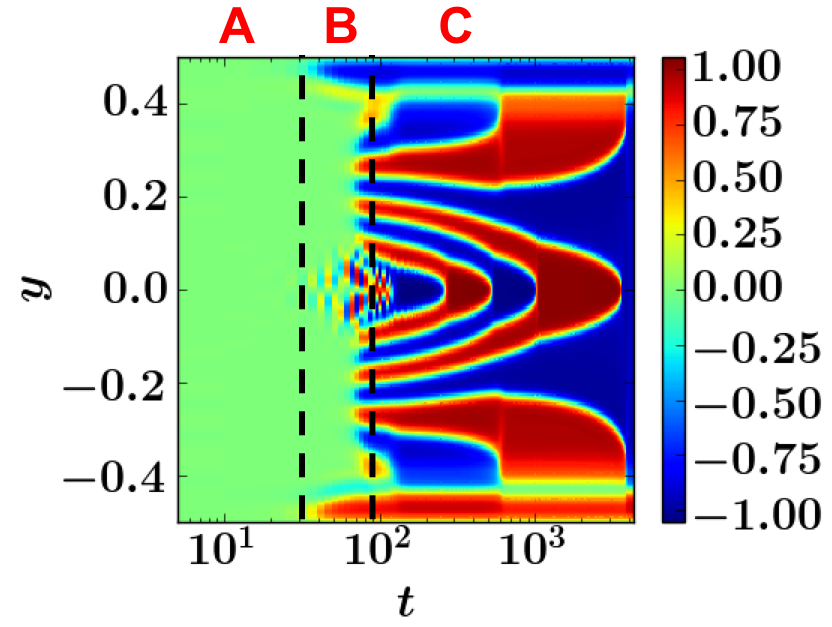} % requires the graphicx package
   \caption{The evolution of $\psi$ at $x=0$ with time (Run2). The three stages are distiguished by black dashed lines, and marked as A, B, and C, respectively. In the target pattern stage (C), the merger process is shown as the corner of the ``$>$" shape.}
   \label{evolution2}
\end{figure}

%\begin{figure}[htbp]
%   \centering
%   \includegraphics[width=0.8\columnwidth]{staircase.png} % requires the graphicx package
%   \caption{The staircase and step merger in confined plasma turbulence: contour plot of the time evolution of $|\nabla n|$ along the plasma radius. Different stages of evolution are: (a) Fast merger of micro-steps and formation of meso-steps. (b) Coalescence of meso-steps to barriers. (c) Barriers propagate along the gradient, condense at boundaries. (d) Stationary profile. Reprinted with permission from Ashourvan and Diamond, Phys. Rev. E 94, 051202(R) (2016). Copyright 2016 American Physical Society. \cite{ashourvan_how_2016}}
%   \label{staircase}
%\end{figure}

\begin{figure}[htbp]
   \centering
   \includegraphics[width=0.8\columnwidth]{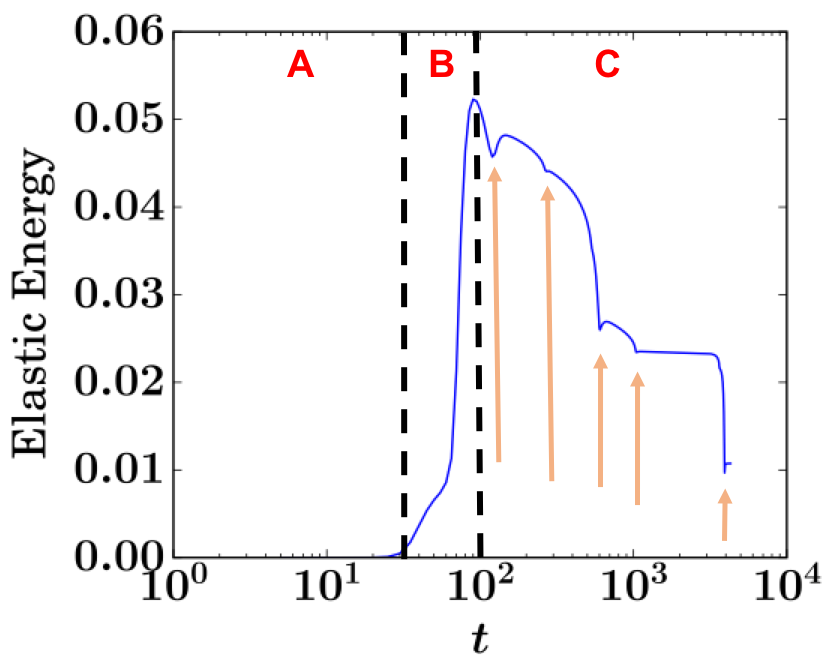} % requires the graphicx package
   \caption{The time evolution of elastic energy (Run2). Note that logarithm scale is used for the $t$ axis. A: the `jelly roll' stage; B: the topological evolution stage; C: the target pattern stage. The dips marked by orange arrows are due to band mergers.}
   \label{surface_tension_single_run}
\end{figure}

The target patterns are metastable. They persist on time scales that are exponentially long relative to the eddy turnover time. During their life, the bands merge with each other very slowly, and the number of bands tends to decrease over time (see Fig. \ref{psi_fields} (f) and (g)). The merger time scales will be discussed in more detail below, after Fig.~\ref{surface_tension_single_run}. Fig.~\ref{evolution2} shows how the concentration field along the $y$ axis at $x=0$ evolves with time. The merger process is shown as the corner of the ``$>$" shape in the plot. The band merger process is similar to step merger in drift-ZF staircases \cite{ashourvan_how_2016,ashourvan_emergence_2017}. The formation and coalescence of meso-steps is analogous to the formation and merger of the target bands in the Cahn-Hilliard system. Because the Cahn-Hilliard system does not support a selected direction, it does not exhibit barrier propagation, as seen in models of drift-ZF staircases.

Fig.~\ref{psi_fields} (h) shows the final steady state. It resembles what is observed during the homogenization of magnetic potential $A$ in MHD. The major difference is that the concentration field is $\psi\sim-1$ in the center of the eddy for the Cahn-Hilliard system, instead of $A\sim0$ in MHD. This result implies that the concentration field is not conserved in our simulation, i.e., the red fluid is lost. This is acceptable, because the Dirichlet boundary conditions do not forbid the transport of matter in or out of the eddy. The choice of this boundary condition is to allow comparisons to studies of flux expulsion.

%\section{Time Evolution of the Elastic Energy}

In our previous study \cite{fan_cascades_2016}, we stressed the analogy between 2D CHNS and 2D MHD. The energy in MHD consists of two parts: kinetic energy and magnetic energy. Similarly, we can also define the energy in the CHNS system to be the sum of kinetic energy $E_K$ and elastic energy $E_B$ \cite{fan_cascades_2016}:
\begin{align}
E_K&=\int\frac{1}{2}\mathbf{v}^2\intd^2x\\
E_B&=\int\frac{1}{2}\xi^2(\nabla\psi)^2\intd^2x
\end{align}
Note that this definition of energy differs from the energy commonly used in studies on the CHNS system ($E=\int[-\frac{1}{2}\psi^2+\frac{1}{4}\psi^4+\frac{1}{2}\xi^2(\nabla\psi)^2+\frac{1}{2}\mathbf{v}^2]\intd^2x$), but it makes it easier to compare to MHD studies. Since the velocity does not change with time in this study, the kinetic energy stays constant. The elastic energy $E_B$ is the analogue of the magnetic energy in MHD. The time evolution of elastic energy for a typical run is shown in Fig.~\ref{surface_tension_single_run}. The 3 stages in the evolution are marked as A, B, and C, respectively.

In the `jelly roll' pattern stage, the elastic energy increases but remains small compared to later stages. In the topology change stage, the elastic energy rises quickly and reaches a maximum value when the topology change is complete. Then, in the target pattern stage, the elastic energy decreases slowly and episodically. When band merger occurs, it appears as a local dip in the elastic energy time evolution plot. Examples of the dips are marked by orange arrows in Fig.~\ref{surface_tension_single_run}. The time scales for band mergers can be obtained in the plot by measuring the time interval of the dips. The time scales for mergers are observed to be linear on the plot with a logarithm scale, so they are exponential on a linear scale. Note that in our normalization, the eddy turnover time is $1$. The band merger time scales differ for each occurrence, but they all are exponentially long relative to the eddy turnover time.

%\section{The effects of $D$ and $\xi$}
In order to observe the phenomena presented above, there is a necessary range of parameters. $\mathrm{Ch}$ should be small, so long as the interface width is resolved ($\xi>h_0$, where $h_0$ is the mesh size). This is because we are interested in cases where the interface width is small compared to the system size. $\mathrm{Pe}$ should be large, so long as the cell's boundary layer is resolved ($L_{BL}>h_0$, where $L_{BL}$ is the width of the boundary layer in $\psi$ at the inner edge of the cell). In MHD, one expects to observe expulsion phenomenon when the magnetic Reynolds number $\mathrm{Rm}$ is large. Analogously, in the Cahn-Hilliard system, we are interested in the large $\mathrm{Pe}$ regime. Similar to the MHD case for which $L_{BL}\sim\mathrm{Rm}^{-1/3}L_0$, the width of the boundary layer in the Cahn-Hilliard system is estimated to be $L_{BL}\sim\mathrm{Pe}^{-1/5}\mathrm{Ch}^{3/5}L_0$. Thus the condition for resolution of $L_{BL}$ is $\mathrm{Pe}^{1/5}\mathrm{Ch}^{-3/5}(\frac{h_0}{L_0})<1$. This expression is obtained by calculating the mixing time scale of the shear + dissipation hybrid case $t_{mix}^{-1}\sim\mathrm{Pe}^{-1/5}\mathrm{Ch}^{2/5}t_0^{-1}$. Note that in the Cahn-Hilliard case, the dissipation is the hyper-diffusion, and the ratio of convection to hyper-diffusion is $\mathrm{Pe}/\mathrm{Ch}^2$.

\begin{figure}[htbp]
   \centering
   \includegraphics[width=\columnwidth]{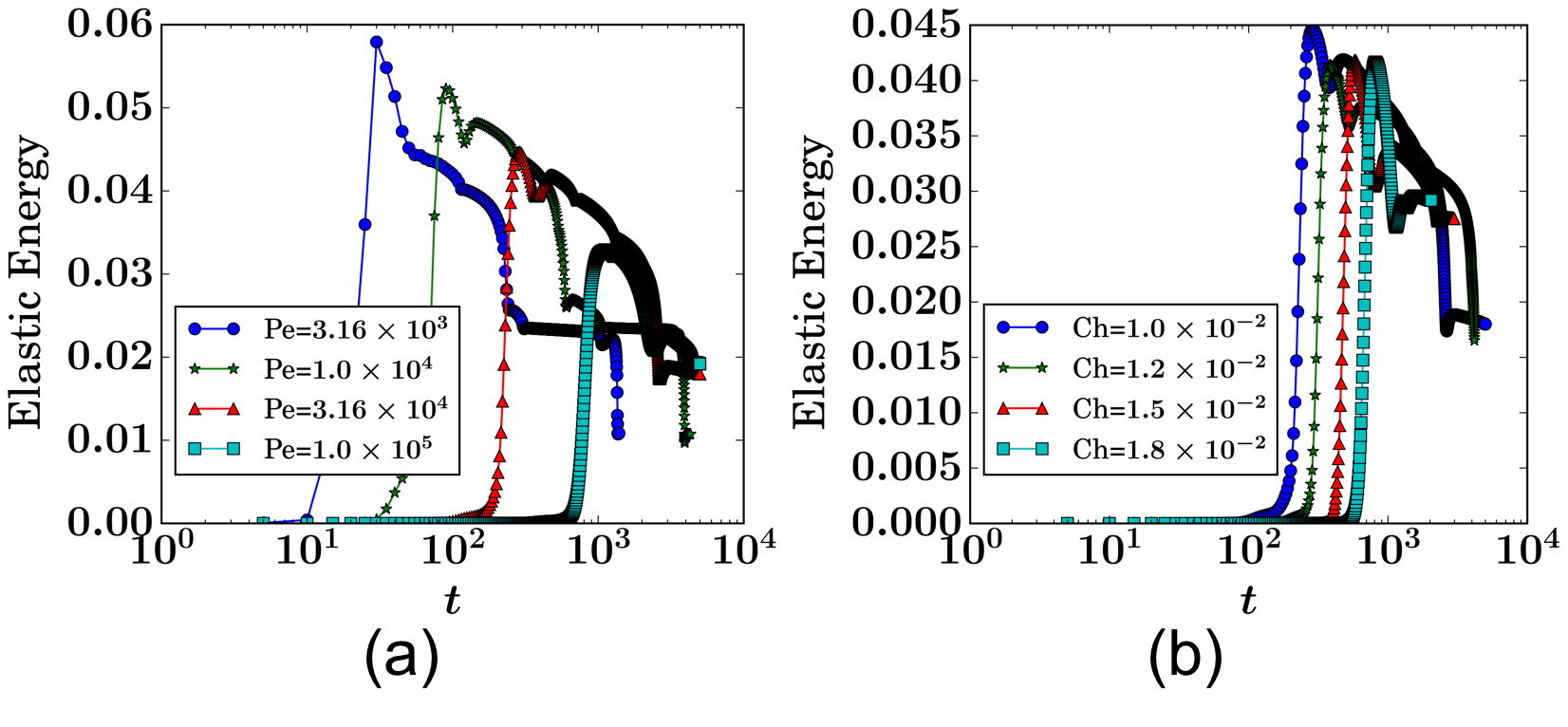} % requires the graphicx package
   \caption{The time evolution of elastic energy for a range of $\mathrm{Pe}$ (a) and $\mathrm{Ch}$ (b).}
   \label{Elastic_Energy}
\end{figure}

\begin{figure}[htbp]
   \centering
   \includegraphics[width=\columnwidth]{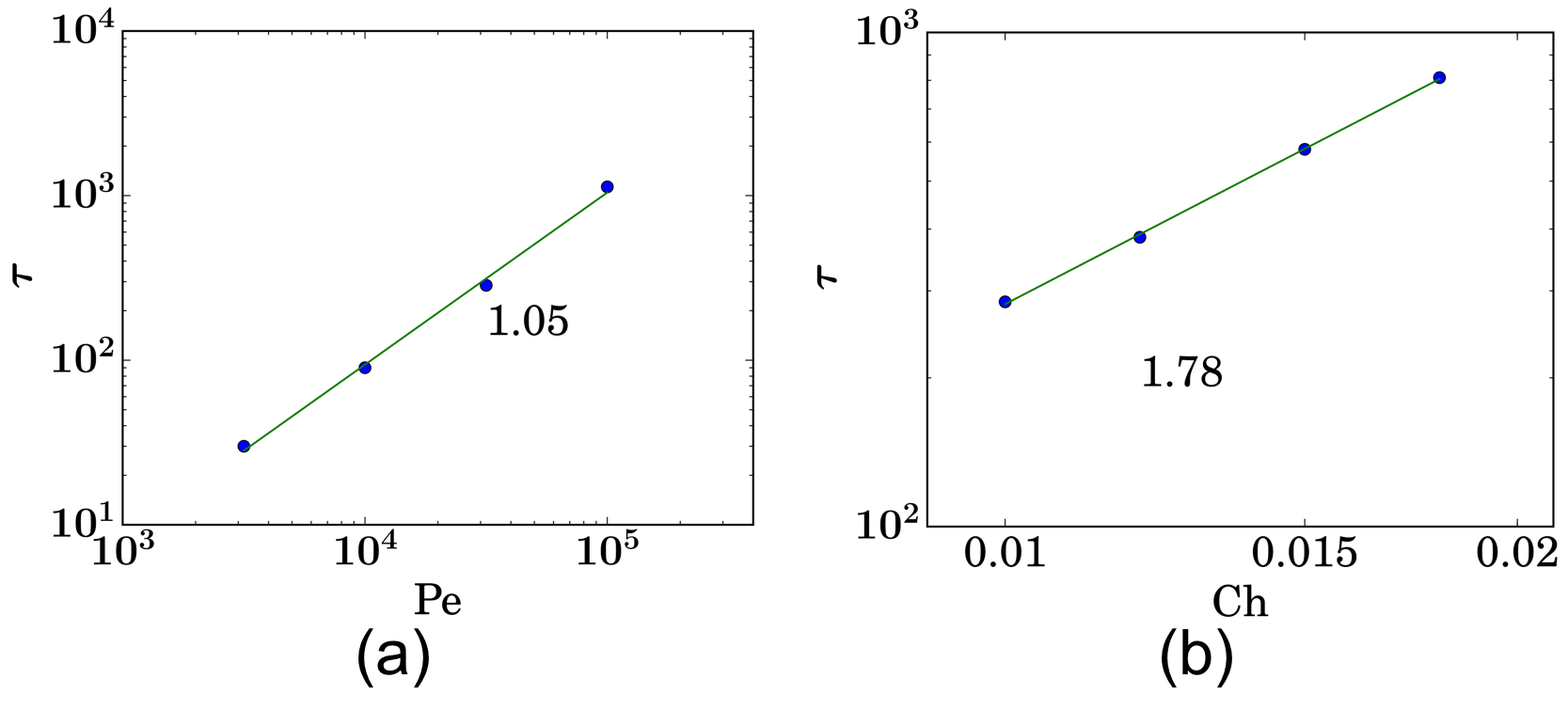} % requires the graphicx package
   \caption{The relationship between the time to reach the maximum elastic energy $\tau$ and the dimensionless parameters $\mathrm{Pe}$ (a) and $\mathrm{Ch}$ (b).}
   \label{tau}
\end{figure}

A parameter scan of $\mathrm{Pe}$ and $\mathrm{Ch}$ is shown in Fig.~\ref{Elastic_Energy}. The elastic energy evolution exhibits the same trend discussed above, and all values of $\mathrm{Pe}$ and $\mathrm{Ch}$ pass through the same three stages. A characteristic time scale in this system is the time to reach the maximum elastic energy $\tau$. $\tau$ is shown to scale as $\tau\sim \mathrm{Pe}^{1.05\pm0.05}\mathrm{Ch}^{1.78\pm0.04}$ (Fig.~\ref{tau}). The error bars reflect only the standard deviations of the linear fits, and the errors from $\tau$ itself are not considered. This relationship can be approximately understood by dimensional analysis: $\tau\propto\xi^2/D$.

%\section{Conclusion and Discussion}
In summary, we have investigated the evolution of the concentration field of the Cahn-Hilliard system in the background of a single convective eddy, motivated by the analogy between CHNS and MHD. This study is an extension of the classic study of flux expulsion in 2D MHD and homogenization of potential vorticity in 2D fluids. We find there are three stages of the evolution: the `jelly roll' pattern stage, the stage of topological evolution, and the target pattern stage. The target bands are metastable: they merge with each other on a time scale exponentially long relative to the eddy turnover time. Band merger occurrences are associated with dips in the elastic energy evolution. The time scale for target pattern evolution and band merger is extended by the imposed shear flow of the eddy. Such flows slow down the merger of bands that is known to occur in, and is natural to, the Cahn-Hilliard system. The band merger process is similar to the drift-ZF staircases in the confined plasma turbulence. The major difference from flux expulsion in MHD is the metastable target pattern stage before reaching the steady homogenized state. Compared to the homogenization of the potential vorticity system, the evolution of the passive scalar in the Cahn-Hilliard system contains additional multi-stage physics, and it exhibits richer dynamics on the long time scale. We also found the time to reach the maximum elastic energy $\tau$ is $\tau\sim\mathrm{Pe}^{1.05\pm0.05}\mathrm{Ch}^{1.78\pm0.04}$. In future work, we will investigate the evolution of the concentration field with the back reaction on the fluid dynamics.

Acknowledgements: We thank David W. Hughes for stimulating discussions and Los Alamos National Laboratory for its hospitality and help with computing resources. This research was supported by the US Department of Energy, Office of Science, Office of Fusion Energy Sciences, under Award No. DE-FG02-04ER54738 and CMTFO Award No. DE-SC0008378.

\bibliography{target_report}

%merlin.mbs apsrev4-1.bst 2010-07-25 4.21a (PWD, AO, DPC) hacked
%Control: key (0)
%Control: author (8) initials jnrlst
%Control: editor formatted (1) identically to author
%Control: production of article title (-1) disabled
%Control: page (0) single
%Control: year (1) truncated
%Control: production of eprint (0) enabled
\begin{thebibliography}{22}%
\makeatletter
\providecommand \@ifxundefined [1]{%
 \@ifx{#1\undefined}
}%
\providecommand \@ifnum [1]{%
 \ifnum #1\expandafter \@firstoftwo
 \else \expandafter \@secondoftwo
 \fi
}%
\providecommand \@ifx [1]{%
 \ifx #1\expandafter \@firstoftwo
 \else \expandafter \@secondoftwo
 \fi
}%
\providecommand \natexlab [1]{#1}%
\providecommand \enquote  [1]{``#1''}%
\providecommand \bibnamefont  [1]{#1}%
\providecommand \bibfnamefont [1]{#1}%
\providecommand \citenamefont [1]{#1}%
\providecommand \href@noop [0]{\@secondoftwo}%
\providecommand \href [0]{\begingroup \@sanitize@url \@href}%
\providecommand \@href[1]{\@@startlink{#1}\@@href}%
\providecommand \@@href[1]{\endgroup#1\@@endlink}%
\providecommand \@sanitize@url [0]{\catcode `\\12\catcode `\$12\catcode
  `\&12\catcode `\#12\catcode `\^12\catcode `\_12\catcode `\%12\relax}%
\providecommand \@@startlink[1]{}%
\providecommand \@@endlink[0]{}%
\providecommand \url  [0]{\begingroup\@sanitize@url \@url }%
\providecommand \@url [1]{\endgroup\@href {#1}{\urlprefix }}%
\providecommand \urlprefix  [0]{URL }%
\providecommand \Eprint [0]{\href }%
\providecommand \doibase [0]{http://dx.doi.org/}%
\providecommand \selectlanguage [0]{\@gobble}%
\providecommand \bibinfo  [0]{\@secondoftwo}%
\providecommand \bibfield  [0]{\@secondoftwo}%
\providecommand \translation [1]{[#1]}%
\providecommand \BibitemOpen [0]{}%
\providecommand \bibitemStop [0]{}%
\providecommand \bibitemNoStop [0]{.\EOS\space}%
\providecommand \EOS [0]{\spacefactor3000\relax}%
\providecommand \BibitemShut  [1]{\csname bibitem#1\endcsname}%
\let\auto@bib@innerbib\@empty
%</preamble>
\bibitem [{\citenamefont {Cahn}\ and\ \citenamefont
  {Hilliard}(1958)}]{cahn_free_1958}%
  \BibitemOpen
  \bibfield  {author} {\bibinfo {author} {\bibfnamefont {J.~W.}\ \bibnamefont
  {Cahn}}\ and\ \bibinfo {author} {\bibfnamefont {J.~E.}\ \bibnamefont
  {Hilliard}},\ }\href {\doibase 10.1063/1.1744102} {\bibfield  {journal}
  {\bibinfo  {journal} {The Journal of Chemical Physics}\ }\textbf {\bibinfo
  {volume} {28}},\ \bibinfo {pages} {258} (\bibinfo {year} {1958})}\BibitemShut
  {NoStop}%
\bibitem [{\citenamefont {Pal}\ \emph {et~al.}(2016)\citenamefont {Pal},
  \citenamefont {Perlekar}, \citenamefont {Gupta},\ and\ \citenamefont
  {Pandit}}]{pal_binary-fluid_2016}%
  \BibitemOpen
  \bibfield  {author} {\bibinfo {author} {\bibfnamefont {N.}~\bibnamefont
  {Pal}}, \bibinfo {author} {\bibfnamefont {P.}~\bibnamefont {Perlekar}},
  \bibinfo {author} {\bibfnamefont {A.}~\bibnamefont {Gupta}}, \ and\ \bibinfo
  {author} {\bibfnamefont {R.}~\bibnamefont {Pandit}},\ }\href {\doibase
  10.1103/PhysRevE.93.063115} {\bibfield  {journal} {\bibinfo  {journal}
  {Physical Review E}\ }\textbf {\bibinfo {volume} {93}},\ \bibinfo {pages}
  {063115} (\bibinfo {year} {2016})}\BibitemShut {NoStop}%
\bibitem [{\citenamefont {Perlekar}\ \emph {et~al.}(2017)\citenamefont
  {Perlekar}, \citenamefont {Pal},\ and\ \citenamefont
  {Pandit}}]{perlekar_two-dimensional_2017}%
  \BibitemOpen
  \bibfield  {author} {\bibinfo {author} {\bibfnamefont {P.}~\bibnamefont
  {Perlekar}}, \bibinfo {author} {\bibfnamefont {N.}~\bibnamefont {Pal}}, \
  and\ \bibinfo {author} {\bibfnamefont {R.}~\bibnamefont {Pandit}},\ }\href
  {\doibase 10.1038/srep44589} {\bibfield  {journal} {\bibinfo  {journal}
  {Scientific Reports}\ }\textbf {\bibinfo {volume} {7}},\ \bibinfo {pages}
  {44589} (\bibinfo {year} {2017})}\BibitemShut {NoStop}%
\bibitem [{\citenamefont {Perlekar}\ \emph {et~al.}(2012)\citenamefont
  {Perlekar}, \citenamefont {Biferale}, \citenamefont {Sbragaglia},
  \citenamefont {Srivastava},\ and\ \citenamefont
  {Toschi}}]{perlekar_droplet_2012}%
  \BibitemOpen
  \bibfield  {author} {\bibinfo {author} {\bibfnamefont {P.}~\bibnamefont
  {Perlekar}}, \bibinfo {author} {\bibfnamefont {L.}~\bibnamefont {Biferale}},
  \bibinfo {author} {\bibfnamefont {M.}~\bibnamefont {Sbragaglia}}, \bibinfo
  {author} {\bibfnamefont {S.}~\bibnamefont {Srivastava}}, \ and\ \bibinfo
  {author} {\bibfnamefont {F.}~\bibnamefont {Toschi}},\ }\href {\doibase
  10.1063/1.4719144} {\bibfield  {journal} {\bibinfo  {journal} {Physics of
  Fluids}\ }\textbf {\bibinfo {volume} {24}},\ \bibinfo {pages} {065101}
  (\bibinfo {year} {2012})}\BibitemShut {NoStop}%
\bibitem [{\citenamefont {Perlekar}\ \emph {et~al.}(2014)\citenamefont
  {Perlekar}, \citenamefont {Benzi}, \citenamefont {Clercx}, \citenamefont
  {Nelson},\ and\ \citenamefont {Toschi}}]{perlekar_spinodal_2014}%
  \BibitemOpen
  \bibfield  {author} {\bibinfo {author} {\bibfnamefont {P.}~\bibnamefont
  {Perlekar}}, \bibinfo {author} {\bibfnamefont {R.}~\bibnamefont {Benzi}},
  \bibinfo {author} {\bibfnamefont {H.~J.}\ \bibnamefont {Clercx}}, \bibinfo
  {author} {\bibfnamefont {D.~R.}\ \bibnamefont {Nelson}}, \ and\ \bibinfo
  {author} {\bibfnamefont {F.}~\bibnamefont {Toschi}},\ }\href
  {http://link.aps.org/doi/10.1103/PhysRevLett.112.014502} {\bibfield
  {journal} {\bibinfo  {journal} {Physical Review Letters}\ }\textbf {\bibinfo
  {volume} {112}} (\bibinfo {year} {2014})}\BibitemShut {NoStop}%
\bibitem [{\citenamefont {Datt}\ \emph {et~al.}(2015)\citenamefont {Datt},
  \citenamefont {Thampi},\ and\ \citenamefont
  {Govindarajan}}]{datt_morphological_2015}%
  \BibitemOpen
  \bibfield  {author} {\bibinfo {author} {\bibfnamefont {C.}~\bibnamefont
  {Datt}}, \bibinfo {author} {\bibfnamefont {S.~P.}\ \bibnamefont {Thampi}}, \
  and\ \bibinfo {author} {\bibfnamefont {R.}~\bibnamefont {Govindarajan}},\
  }\href {\doibase 10.1103/PhysRevE.91.010101} {\bibfield  {journal} {\bibinfo
  {journal} {Physical Review E}\ }\textbf {\bibinfo {volume} {91}},\ \bibinfo
  {pages} {010101} (\bibinfo {year} {2015})}\BibitemShut {NoStop}%
\bibitem [{\citenamefont {Tierra}\ and\ \citenamefont
  {Guill\'en-Gonz\'alez}(2014)}]{tierra_numerical_2014}%
  \BibitemOpen
  \bibfield  {author} {\bibinfo {author} {\bibfnamefont {G.}~\bibnamefont
  {Tierra}}\ and\ \bibinfo {author} {\bibfnamefont {F.}~\bibnamefont
  {Guill\'en-Gonz\'alez}},\ }\href {\doibase 10.1007/s11831-014-9112-1}
  {\bibfield  {journal} {\bibinfo  {journal} {Archives of Computational Methods
  in Engineering}\ ,\ \bibinfo {pages} {1}} (\bibinfo {year}
  {2014})}\BibitemShut {NoStop}%
\bibitem [{\citenamefont {Guill\'en-Gonz\'alez}\ and\ \citenamefont
  {Tierra}(2014)}]{guillen-gonzalez_second_2014}%
  \BibitemOpen
  \bibfield  {author} {\bibinfo {author} {\bibfnamefont {F.}~\bibnamefont
  {Guill\'en-Gonz\'alez}}\ and\ \bibinfo {author} {\bibfnamefont
  {G.}~\bibnamefont {Tierra}},\ }\href {\doibase 10.1016/j.camwa.2014.07.014}
  {\bibfield  {journal} {\bibinfo  {journal} {Computers \& Mathematics with
  Applications}\ }\textbf {\bibinfo {volume} {68}},\ \bibinfo {pages} {821}
  (\bibinfo {year} {2014})}\BibitemShut {NoStop}%
\bibitem [{\citenamefont {Fan}\ \emph {et~al.}(2016)\citenamefont {Fan},
  \citenamefont {Diamond}, \citenamefont {Chac\'on},\ and\ \citenamefont
  {Li}}]{fan_cascades_2016}%
  \BibitemOpen
  \bibfield  {author} {\bibinfo {author} {\bibfnamefont {X.}~\bibnamefont
  {Fan}}, \bibinfo {author} {\bibfnamefont {P.~H.}\ \bibnamefont {Diamond}},
  \bibinfo {author} {\bibfnamefont {L.}~\bibnamefont {Chac\'on}}, \ and\
  \bibinfo {author} {\bibfnamefont {H.}~\bibnamefont {Li}},\ }\href {\doibase
  10.1103/PhysRevFluids.1.054403} {\bibfield  {journal} {\bibinfo  {journal}
  {Physical Review Fluids}\ }\textbf {\bibinfo {volume} {1}},\ \bibinfo {pages}
  {054403} (\bibinfo {year} {2016})}\BibitemShut {NoStop}%
\bibitem [{\citenamefont {Ruiz}\ and\ \citenamefont
  {Nelson}(1981)}]{ruiz_turbulence_1981}%
  \BibitemOpen
  \bibfield  {author} {\bibinfo {author} {\bibfnamefont {R.}~\bibnamefont
  {Ruiz}}\ and\ \bibinfo {author} {\bibfnamefont {D.~R.}\ \bibnamefont
  {Nelson}},\ }\href
  {http://journals.aps.org/pra/abstract/10.1103/PhysRevA.23.3224} {\bibfield
  {journal} {\bibinfo  {journal} {Physical Review A}\ }\textbf {\bibinfo
  {volume} {23}},\ \bibinfo {pages} {3224} (\bibinfo {year}
  {1981})}\BibitemShut {NoStop}%
\bibitem [{\citenamefont {Weiss}(1966)}]{weiss_expulsion_1966}%
  \BibitemOpen
  \bibfield  {author} {\bibinfo {author} {\bibfnamefont {N.~O.}\ \bibnamefont
  {Weiss}},\ }\href {\doibase 10.1098/rspa.1966.0173} {\bibfield  {journal}
  {\bibinfo  {journal} {Proceedings of the Royal Society of London A:
  Mathematical, Physical and Engineering Sciences}\ }\textbf {\bibinfo {volume}
  {293}},\ \bibinfo {pages} {310} (\bibinfo {year} {1966})}\BibitemShut
  {NoStop}%
\bibitem [{\citenamefont {Gilbert}\ \emph {et~al.}(2016)\citenamefont
  {Gilbert}, \citenamefont {Mason},\ and\ \citenamefont
  {Tobias}}]{gilbert_flux_2016}%
  \BibitemOpen
  \bibfield  {author} {\bibinfo {author} {\bibfnamefont {A.~D.}\ \bibnamefont
  {Gilbert}}, \bibinfo {author} {\bibfnamefont {J.}~\bibnamefont {Mason}}, \
  and\ \bibinfo {author} {\bibfnamefont {S.~M.}\ \bibnamefont {Tobias}},\
  }\href
  {https://www.cambridge.org/core/journals/journal-of-fluid-mechanics/article/flux-expulsion-with-dynamics/0936A7CDF7F23929806AF926681780EC}
  {\bibfield  {journal} {\bibinfo  {journal} {Journal of Fluid Mechanics}\
  }\textbf {\bibinfo {volume} {791}},\ \bibinfo {pages} {568} (\bibinfo {year}
  {2016})}\BibitemShut {NoStop}%
\bibitem [{\citenamefont {Moffatt}(1983)}]{moffatt_magnetic_1983}%
  \BibitemOpen
  \bibfield  {author} {\bibinfo {author} {\bibfnamefont {H.~K.}\ \bibnamefont
  {Moffatt}},\ }\href@noop {} {\emph {\bibinfo {title} {Magnetic field
  generation in electrically conducting fluids}}}\ (\bibinfo  {publisher}
  {Cambridge University Press},\ \bibinfo {year} {1983})\BibitemShut {NoStop}%
\bibitem [{\citenamefont {Rhines}\ and\ \citenamefont
  {Young}(1982)}]{rhines_homogenization_1982}%
  \BibitemOpen
  \bibfield  {author} {\bibinfo {author} {\bibfnamefont {P.~B.}\ \bibnamefont
  {Rhines}}\ and\ \bibinfo {author} {\bibfnamefont {W.~R.}\ \bibnamefont
  {Young}},\ }\href {\doibase 10.1017/S0022112082002250} {\bibfield  {journal}
  {\bibinfo  {journal} {Journal of Fluid Mechanics}\ }\textbf {\bibinfo
  {volume} {122}},\ \bibinfo {pages} {347} (\bibinfo {year}
  {1982})}\BibitemShut {NoStop}%
\bibitem [{\citenamefont {Rhines}\ and\ \citenamefont
  {Young}(1983)}]{rhines_how_1983}%
  \BibitemOpen
  \bibfield  {author} {\bibinfo {author} {\bibfnamefont {P.~B.}\ \bibnamefont
  {Rhines}}\ and\ \bibinfo {author} {\bibfnamefont {W.~R.}\ \bibnamefont
  {Young}},\ }\href {http://journals.cambridge.org/article_S0022112083001822}
  {\bibfield  {journal} {\bibinfo  {journal} {Journal of Fluid Mechanics}\
  }\textbf {\bibinfo {volume} {133}},\ \bibinfo {pages} {133} (\bibinfo {year}
  {1983})}\BibitemShut {NoStop}%
\bibitem [{\citenamefont {Chac\'on}\ \emph {et~al.}(2002)\citenamefont
  {Chac\'on}, \citenamefont {Knoll},\ and\ \citenamefont
  {Finn}}]{chacon_implicit_2002}%
  \BibitemOpen
  \bibfield  {author} {\bibinfo {author} {\bibfnamefont {L.}~\bibnamefont
  {Chac\'on}}, \bibinfo {author} {\bibfnamefont {D.}~\bibnamefont {Knoll}}, \
  and\ \bibinfo {author} {\bibfnamefont {J.}~\bibnamefont {Finn}},\ }\href
  {\doibase 10.1006/jcph.2002.7015} {\bibfield  {journal} {\bibinfo  {journal}
  {Journal of Computational Physics}\ }\textbf {\bibinfo {volume} {178}},\
  \bibinfo {pages} {15} (\bibinfo {year} {2002})}\BibitemShut {NoStop}%
\bibitem [{\citenamefont {Chac\'on}\ \emph {et~al.}(2003)\citenamefont
  {Chac\'on}, \citenamefont {Knoll},\ and\ \citenamefont
  {Finn}}]{chacon_hall_2003}%
  \BibitemOpen
  \bibfield  {author} {\bibinfo {author} {\bibfnamefont {L.}~\bibnamefont
  {Chac\'on}}, \bibinfo {author} {\bibfnamefont {D.}~\bibnamefont {Knoll}}, \
  and\ \bibinfo {author} {\bibfnamefont {J.}~\bibnamefont {Finn}},\ }\href
  {\doibase 10.1016/S0375-9601(02)01807-8} {\bibfield  {journal} {\bibinfo
  {journal} {Physics Letters A}\ }\textbf {\bibinfo {volume} {308}},\ \bibinfo
  {pages} {187} (\bibinfo {year} {2003})}\BibitemShut {NoStop}%
\bibitem [{\citenamefont {Liu}\ \emph {et~al.}(2013)\citenamefont {Liu},
  \citenamefont {Ded\`e}, \citenamefont {Evans}, \citenamefont {Borden},\ and\
  \citenamefont {Hughes}}]{liu_isogeometric_2013}%
  \BibitemOpen
  \bibfield  {author} {\bibinfo {author} {\bibfnamefont {J.}~\bibnamefont
  {Liu}}, \bibinfo {author} {\bibfnamefont {L.}~\bibnamefont {Ded\`e}},
  \bibinfo {author} {\bibfnamefont {J.~A.}\ \bibnamefont {Evans}}, \bibinfo
  {author} {\bibfnamefont {M.~J.}\ \bibnamefont {Borden}}, \ and\ \bibinfo
  {author} {\bibfnamefont {T.~J.~R.}\ \bibnamefont {Hughes}},\ }\href {\doibase
  10.1016/j.jcp.2013.02.008} {\bibfield  {journal} {\bibinfo  {journal}
  {Journal of Computational Physics}\ }\textbf {\bibinfo {volume} {242}},\
  \bibinfo {pages} {321} (\bibinfo {year} {2013})}\BibitemShut {NoStop}%
\bibitem [{\citenamefont {Hashimoto}\ \emph {et~al.}(1993)\citenamefont
  {Hashimoto}, \citenamefont {Takebe},\ and\ \citenamefont
  {Asakawa}}]{hashimoto_phase_1993}%
  \BibitemOpen
  \bibfield  {author} {\bibinfo {author} {\bibfnamefont {T.}~\bibnamefont
  {Hashimoto}}, \bibinfo {author} {\bibfnamefont {T.}~\bibnamefont {Takebe}}, \
  and\ \bibinfo {author} {\bibfnamefont {K.}~\bibnamefont {Asakawa}},\ }\href
  {\doibase 10.1016/0378-4371(93)90367-D} {\bibfield  {journal} {\bibinfo
  {journal} {Physica A: Statistical Mechanics and its Applications}\ }\textbf
  {\bibinfo {volume} {194}},\ \bibinfo {pages} {338} (\bibinfo {year}
  {1993})}\BibitemShut {NoStop}%
\bibitem [{\citenamefont {Hashimoto}\ \emph {et~al.}(1995)\citenamefont
  {Hashimoto}, \citenamefont {Matsuzaka}, \citenamefont {Moses},\ and\
  \citenamefont {Onuki}}]{hashimoto_string_1995}%
  \BibitemOpen
  \bibfield  {author} {\bibinfo {author} {\bibfnamefont {T.}~\bibnamefont
  {Hashimoto}}, \bibinfo {author} {\bibfnamefont {K.}~\bibnamefont
  {Matsuzaka}}, \bibinfo {author} {\bibfnamefont {E.}~\bibnamefont {Moses}}, \
  and\ \bibinfo {author} {\bibfnamefont {A.}~\bibnamefont {Onuki}},\ }\href
  {\doibase 10.1103/PhysRevLett.74.126} {\bibfield  {journal} {\bibinfo
  {journal} {Physical Review Letters}\ }\textbf {\bibinfo {volume} {74}},\
  \bibinfo {pages} {126} (\bibinfo {year} {1995})}\BibitemShut {NoStop}%
\bibitem [{\citenamefont {Ashourvan}\ and\ \citenamefont
  {Diamond}(2016)}]{ashourvan_how_2016}%
  \BibitemOpen
  \bibfield  {author} {\bibinfo {author} {\bibfnamefont {A.}~\bibnamefont
  {Ashourvan}}\ and\ \bibinfo {author} {\bibfnamefont {P.~H.}\ \bibnamefont
  {Diamond}},\ }\href {\doibase 10.1103/PhysRevE.94.051202} {\bibfield
  {journal} {\bibinfo  {journal} {Physical Review E}\ }\textbf {\bibinfo
  {volume} {94}},\ \bibinfo {pages} {051202} (\bibinfo {year}
  {2016})}\BibitemShut {NoStop}%
\bibitem [{\citenamefont {Ashourvan}\ and\ \citenamefont
  {Diamond}(2017)}]{ashourvan_emergence_2017}%
  \BibitemOpen
  \bibfield  {author} {\bibinfo {author} {\bibfnamefont {A.}~\bibnamefont
  {Ashourvan}}\ and\ \bibinfo {author} {\bibfnamefont {P.~H.}\ \bibnamefont
  {Diamond}},\ }\href {\doibase 10.1063/1.4973660} {\bibfield  {journal}
  {\bibinfo  {journal} {Physics of Plasmas}\ }\textbf {\bibinfo {volume}
  {24}},\ \bibinfo {pages} {012305} (\bibinfo {year} {2017})}\BibitemShut
  {NoStop}%
\end{thebibliography}%
\nocite{*}

\end{document}